\documentclass{llncs}
\usepackage{color,soul}
\usepackage{lineno,hyperref}
\usepackage{amssymb}
\usepackage{lineno,hyperref}
\usepackage{amsmath}
\usepackage[font=small,skip=0pt]{subcaption}
\usepackage{graphicx}
\usepackage{commath}
\usepackage{tabularx, booktabs}
\usepackage{multirow}
\usepackage[para,online,flushleft]{threeparttable}
\usepackage{rotating}
\usepackage{algorithm}
\usepackage[noend]{algpseudocode}
\usepackage{color, colortbl}
\usepackage[table]{xcolor}%
\definecolor{Gray}{gray}{0.9}
\usepackage{bbding}

\begin{document}

\title{A Multi-task Network to Detect Junctions in Retinal Vasculature}
%
%\titlerunning{Hamiltonian Mechanics}  % abbreviated title (for running head)
%                                     also used for the TOC unless
%                                     \toctitle is used
%
\author{Fatmat\"{u}lzehra Uslu\Envelope\inst{1}\orcidID{0000-0001-7153-7583} \and Anil Anthony Bharath\inst{1}}
%\authorrunning{Fatmatulzehra Uslu \and Anil A. Bharath}
\institute{BICV group, Bioengineering Department, Imperial College London, UK
\email{fzehrauslu@gmail.com}, \email{a.bharath@imperial.ac.uk}}

%\author{Ivar Ekeland\inst{1} \and Roger Temam\inst{2}
%Jeffrey Dean \and David Grove \and Craig Chambers \and Kim~B.~Bruce \and
%Elsa Bertino}
%
%\authorrunning{Ivar Ekeland et al.} % abbreviated author list (for running head)
%
%%%% list of authors for the TOC (use if author list has to be modified)
%\tocauthor{Ivar Ekeland, Roger Temam, Jeffrey Dean, David Grove,
%Craig Chambers, Kim B. Bruce, and Elisa Bertino}
%

\maketitle              

\begin{abstract}
Junctions in the retinal vasculature are key points to be able to extract its topology, but they vary in appearance, depending on vessel density, width and branching/crossing angles. The complexity of junction patterns is usually accompanied by a scarcity of labels, which discourages the usage of very deep networks for their detection. We propose a multi-task network, generating labels for vessel interior, centerline, edges and junction patterns, to provide additional information to facilitate junction detection. After the initial detection of potential junctions in junction-selective probability maps, candidate locations are re-examined in centerline probability maps to verify if they connect at least $3$ branches. The experiments on the DRIVE and IOSTAR showed that our method outperformed a recent study in which a popular deep network was trained as a classifier to find junctions. Moreover, the proposed approach is applicable to unseen datasets with the same degree of success, after training it only once. 
\end{abstract}
\section{Introduction}
The variations on geometry and topology of retinal vasculature can give information about the health of the eye. Junctions in the retinal vasculature represent important landmarks to extract retinal vessel topology, and can also facilitate image registration. The appearance of junctions in fundus images can vary markedly depending on branching angles, vessel widths and junction sizes, somewhat complicating their detection.

Several previous studies have made use of algorithms designed to work on binary vessel maps, usually on vessel skeletons \cite{azzopardi2013automatic,calvo2011automatic,hamad2014automatic,pratt2017automatic}. Generally, these methods relied on examining the intersection numbers in the skeletons \cite{calvo2011automatic} accompanied with vessel widths, lengths and branching angles \cite{hamad2014automatic}. However, the skeletons can misrepresent vessel topology due to either/both gaps in vessels in segmented vessel maps or, missing or false vessels in these maps.   

Recently, deep learning methods have been frequently used for medical image analysis. However, they have been rarely applied to junction detection in retinal vasculature. As far as we are aware, the only study is Pratt \textit{et al}'s \cite{pratt2017automatic}, where a popular network was trained as a classifier to identify image patches with junctions. The image patches contained vessel skeletons, generated from binary vessel maps. The major factor restricting the application of deep architectures on this task seems the shortage of training data.  The number of publicly accessible image datasets is only two \cite{abbasi2016automatic}, as far as we are aware. Also, the task is extremely skewed, where the fraction of junction pixels in a fundus image is lower than $10^{-4}$ \cite{abbasi2016automatic}. These factors -- and primarily the imbalanced nature of the junction presence -- complicate the training of deep networks containing a large number of parameters.

In this study, we introduce a multi-task network for junction detection on fundus images, which can deal with the shortage of labeled data and highly skewed nature of the task. We also present a junction probability map, which significantly facilitates finding junctions by removing possible variations in vessel thickness. Experiments on the DRIVE and IOSTAR show our method outperformed Pratt \textit{et al.}'s method \cite{pratt2017automatic}, despite operating directly on fundus images.

\section{Method \label{sec:Method}}
Our method has three stages: (i) training our multi-task network, (ii) initial junction search and (iii) refined junction search. 
\subsection{Learning Junction Patterns With Multi-task Network}
The amount of labeled data for junction locations in retinal vasculature is not sufficient to successfully train a deep network with many parameters. In contrast, vessel segmentation has been realized by deep networks in increasing numbers of studies \cite{srinidhi2017recent}. In order to deal with the shortage of labeled data for the junction detection, we propose a multi-task network, which simultaneously generates label patches for vessel interior, centerline, edge locations and junction patterns. 

\begin{figure}[!ht]
\centering
{\includegraphics[scale=0.3]{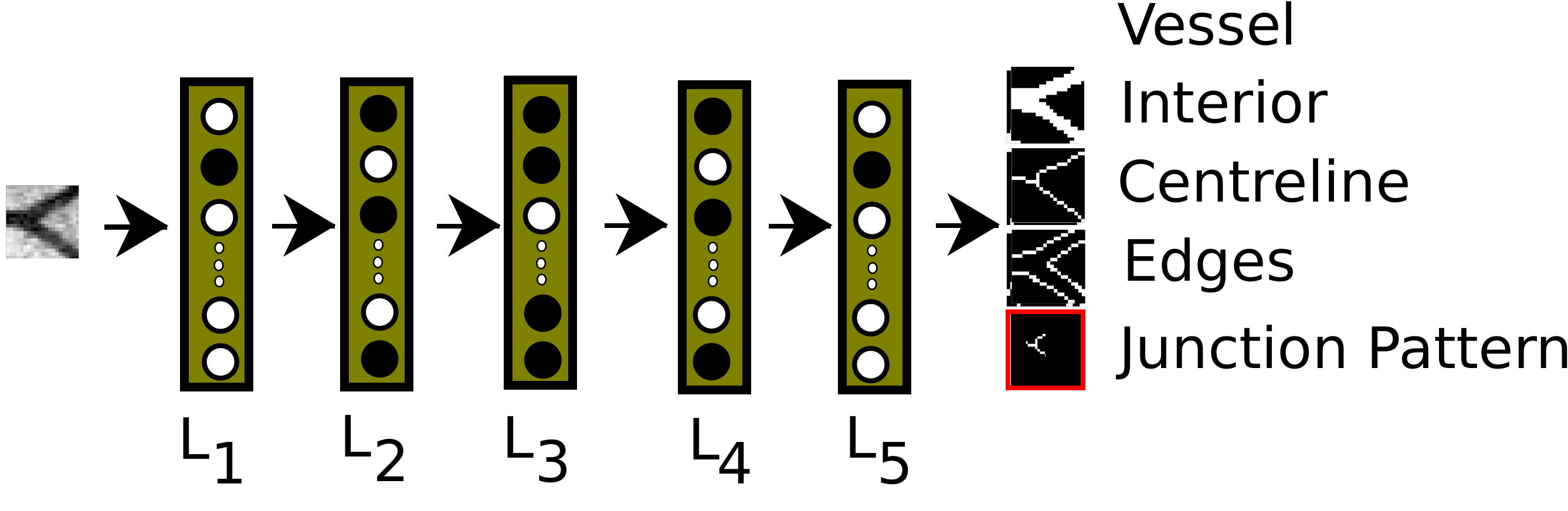}} 
\caption{The generation of junction pattern label patches with a fully connected network with $5$ hidden layers. \label{fig:T-DBNMultilabelling}}
\end{figure}

The proposed network for this task is a fully connected network, which is initialized with weights learned by training a stack of Restricted Boltzmann Machines (RBMs) \cite{hinton2012practical} on five types of image patches: fundus image patches, label patches for vessel interior, centerline, edges and junction patterns. Label patches can be combined, as suggested by Li \textit{et al} \cite{Li2015}, to generate a likelihood map for the whole fundus image (see Fig.~\ref{fig:T-DBNMultilabelling}). After initialization, the network is trained patch-wise with the $l_{2}$ loss function in Eqn.~(\ref{eqn:LossFunc}):
\begin{equation} \label{eqn:LossFunc}
\mathcal{L}=||\mathcal{V}_{i}^{p}-\tilde{\mathcal{V}}_{i}^{p}||_{2} + ||\mathcal{V}_{c}^{p}-\tilde{\mathcal{V}}_{c}^{p}||_{2}+||\mathcal{V}_{e}^{p}-\tilde{\mathcal{V}}_{e}^{p}||_{2}+||\mathcal{V}_{j}^{p}-\tilde{\mathcal{V}}_{j}^{p}||_{2}
\end{equation}
where $\tilde{\mathcal{V}}_{i}^{p}$,$\tilde{\mathcal{V}}_{c}^{p}$, $\tilde{\mathcal{V}}_{e}^{p}$ and $\tilde{\mathcal{V}}_{j}^{p}$ respectively represent estimates of ground truth label patches for vessel interior ($\mathcal{V}_{i}^{p}$), centerline ($\mathcal{V}_{c}^{p}$), edges ($\mathcal{V}_{e}^{p}$) and junction patterns ($\mathcal{V}_{j}^{p}$) by the network.

One key advantage of this network is that -- despite the ability to perform more than one task -- it has relatively fewer parameters, slightly over one million, than those of many state-of-the-art networks. For example, $Res16$ \cite{he2016deep} has over eleven million parameters, which was used by Pratt \textit{et al.} to classify image patches with junctions \cite{pratt2017automatic}. Also, pretraining can provide a good initialization for the network parameters and facilitate training of the network in the face of the scarcity of labeled data. 

\subsection{Initial Junction Search \label{sec:InitialSearch}}
\begin{figure}[!ht]
  \centering
  {\includegraphics[scale=0.45,keepaspectratio,trim=0.5cm 0cm 0.8cm 0.5cm,clip=true]{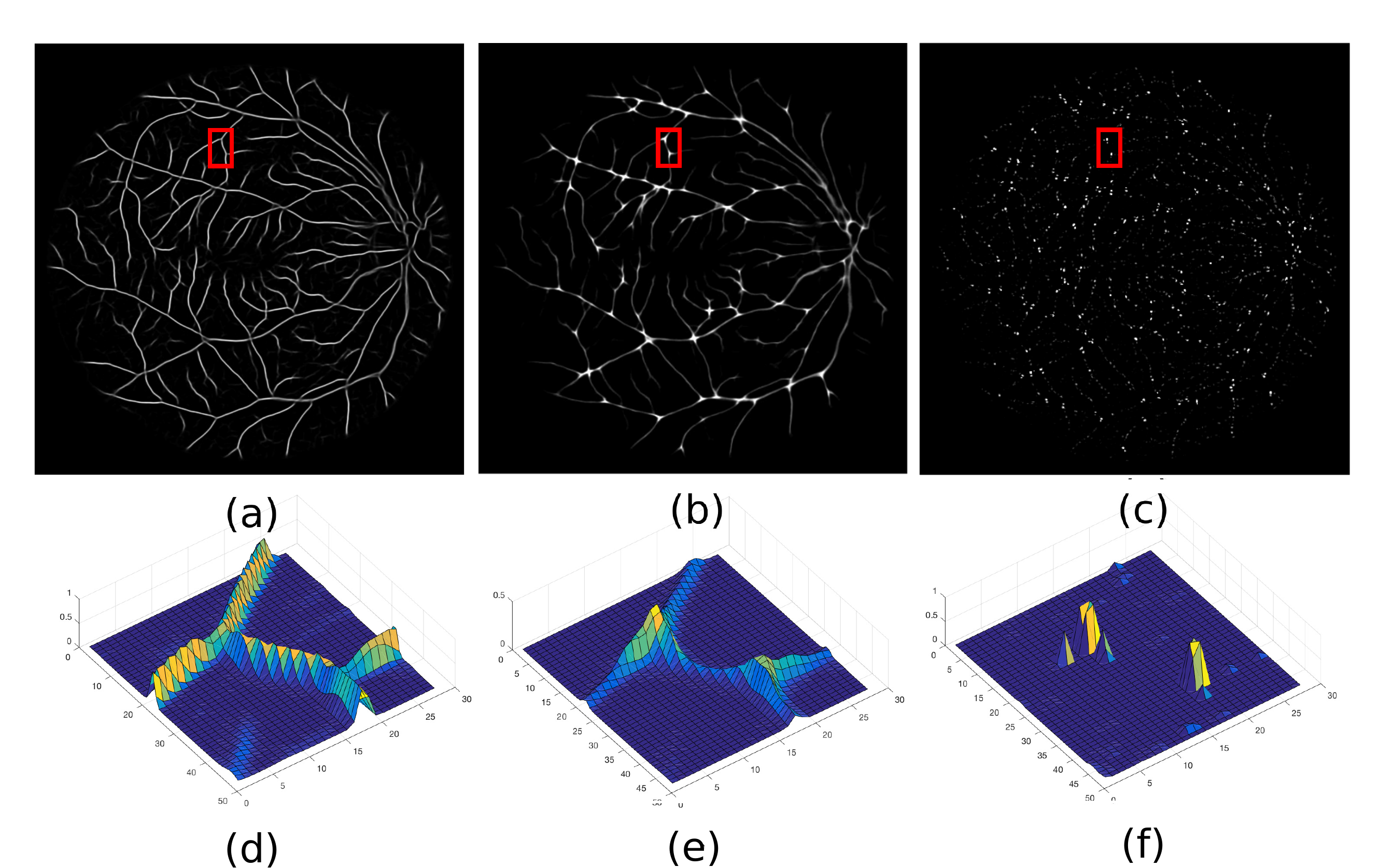}}
  \vskip 0.2cm
\caption{(a) A centerline probability map generated by the proposed method for the $19^{th}$ image in the DRIVE test set. (b) Its junction probability map (c) The map is generated after multiplying eigenvalues calculated on the junction probability map in (b). (d)-(f) $3D$ visualization of regions framed in red in (a)-(c). \label{fig:BifurcationPeak}} 
\end{figure}

The junction probability maps produced by our method, shown in Figs. \ref{fig:BifurcationPeak}(b) and (e), demonstrate significantly larger probabilities in the presence of junctions, but can also yield occasional weak responses to the centrelines of larger vessels. The center regions of the junction patterns on the maps appear to be blob-like and can be detected by eigen-analysis on Hessian decomposition \cite{frangi1998multiscale} of junction probability maps. Bright blob-like structures have negative eigenvalues with similar magnitudes; $| \overrightarrow{E_{1}} | \simeq | \overrightarrow{E_{2}}|$  \cite{frangi1998multiscale}. 
\vspace{-0.2cm}

Because the magnitude of junction probabilities can vary depending on the network's recognition ability for these structures, we normalize eigenvalues locally in the range of $[0,1]$ to make the detection of junction centers more immune to variations on junction probabilities. Simply multiplying normalized eigenvalues for each pixel is observed to generate greater values for junctions than those for vessel centerlines, as demonstrated in Figs. \ref{fig:BifurcationPeak}(c) and (f). This map is thresholded to locate junctions, but yields some false positives. The following section will describe a method to eliminate these false positives. 

\subsection{Refined Junction Search \label{sec:RefinedSearch}}
We count branch numbers on our vessel centerline probability maps, demonstrated in Figs. \ref{fig:BifurcationPeak}(a) and (d), to remove false positives obtained in the initial search. In contrast to previous studies counting intersection numbers in vessel skeletons to identify junction locations \cite{calvo2011automatic}, we use centerline probability maps to avoid mistakes, which can occur due to further processing of segmented vessel maps. For example, skeletonization may lead to false branches \cite{calvo2011automatic}. 

\begin{figure}[!ht]
  \centering
  {\includegraphics[scale=1.5,keepaspectratio,trim=0cm 0cm 1.5cm 0cm,clip=true]{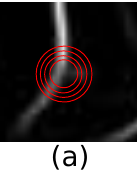}~~~~~~~~~~~~~~~
\includegraphics[scale=1.5,keepaspectratio,trim=1.5cm 0cm 0cm 0cm,clip=true]{INCFIGS/Figures/CirclesAroundJunctions2-eps-converted-to.pdf}}
\vskip 0.2cm
\caption{Two examples from refined junction search: a falsely detected location with two branches in (a) is eliminated and the location with three branches in (b) is kept despite having lower centerline probabilities near the junction center. \label{fig:RefinedSearch}} 
\end{figure}

We locate four circles centered at potential junctions obtained in the previous step, and decide on branch numbers by calculating the most repetitive and largest intersection number over the circles (see Fig. \ref{fig:RefinedSearch}). The reason for using multiple circles is to reduce the possibility of detecting false branches in the centerline probability maps. 

\section{Experimental Setup and Material} 
\paragraph*{Parameter Settings:} \label{NetworkProperties} 

The proposed network consists of the input layer with $256$ units, $5$ hidden layers with $400$ units in each layer and the output layer with $1024$ units. The size of an input image patch and that of a label patch are $16$ by $16$ pixels.

The initialization of network weights in pretraining was made by sampling from a normal distribution $\mathcal{N}(0,0.001)$. After training the network for $50$ epochs with a learning rate of $0.005$, a momentum of $0.5$ for the first $5$ epochs and that of $0.9$ for the rest of pretraining, as recommended in \cite{hinton2012practical}, we finetuned the weights for $120$ epochs with a learning rate of $0.08$. 

The parameters for junction detection (initial and refined search) were as follows. Radii of circles were $3$ to $6$ pixels. Threshold for junction probability map was $0.3$ and that for centerline probability maps to calculate branch number was $0.1$. 

\paragraph*{Performance Evaluation:}
We used precision, recall and $F_{1}$ score for performance evaluation, similar to \cite{abbasi2016automatic,azzopardi2013automatic,zhang2018line}, and accepted junctions estimated in the distance of $5$ pixels to actual junctions true. These metrics are calculated with the following definitions over the entire set of junctions in a dataset: $Recall= \frac{TP}{TP+FN}$, $Precision= \frac{TP}{TP+FP}$ and $F_{1} score= \frac{2\cdot Recall \cdot Precision}{Recall+ Precision}$, where $TP$, $FN$ and $FP$ respectively denote the number of correctly labeled junctions, that of missed junctions and that of mistakenly labeled junctions. 

\paragraph*{Material:}
We evaluated the performance of our method on two fundus image datasets, which are captured by different image modalities. The DRIVE \cite{staal2004ridge} is the most popular dataset for the evaluation of retinal vessel segmentation methods, which contains $40$ images with a resolution of $768 \times584$ pixels, acquired with a nonmydriatic CCD camera. The IOSTAR \cite{abbasi2016automatic} consists of $24$ scanning laser ophthalmoscope (SLO) images. Both datasets have ground truth vessel maps and junction locations \cite{abbasi2016automatic} \footnote{The ground truth data is available at \url{http://retinacheck.org/datasets}}, and are publicly available. The number of junctions per image, including both bifurcations and crossings, is roughly $130$ for the DRIVE and $78$ for IOSTAR.

\paragraph{Training Data Preparation:} We obtained ground truth labels for, respectively, vessel centerline and edges by applying a simple thinning and edge detection method to ground truth vessel maps. Later, we generated junction pattern labels by using ground truth centerline maps: we firstly located masks of $5 \times 5$ pixels centered at ground truth junction locations then removed centerline labels outside these masks. The training dataset contains the last $20$ images in the DRIVE and test set includes the first $20$ images, which are separated by authors in \cite{staal2004ridge}. We prepared $18,000$ mini-batches, each contains $100$ training samples. Because the network was only trained with the DRIVE dataset, there was no need for preparing a training set for IOSTAR.

\section{Results}
We evaluated the performance of the proposed method on the DRIVE test set and the complete set of IOSTAR. Similar to Pratt \textit{et al.}'s training \cite{pratt2017automatic}, we trained our network with only the DRIVE training set. The reason for this is that the number of ground truth junctions per image in IOSTAR is almost half that given for the DRIVE dataset. We also assessed the performance of our method on ground truth vessel maps, to make a fair performance comparison with previous methods that utilize these maps. Table \ref{table:ConfusionMatrixDRIVE_IOSTAR} presents our findings and compares them with those of previously reported methods.  

\begin{table}[!ht]
\centering
\caption{The performance comparison for junction detection.
\label{table:ConfusionMatrixDRIVE_IOSTAR}}
\resizebox{1\textwidth}{!}{%
\begin{threeparttable}
\begin{tabular} {@{} lcccccccc @{} } \toprule %
&& &\multicolumn{3}{c}{ DRIVE}&\multicolumn{3}{c}{ IOSTAR} \\ \cmidrule(lr){4-6} \cmidrule(lr){7-9}
&&Input Image Type&Precision&Recall&$F_{1}$ score&Precision&Recall&$F_{1}$ score\\  \cmidrule(lr){3-3} \cmidrule(lr){4-4}\cmidrule(lr){5-5} \cmidrule(lr){6-6} \cmidrule(lr){7-7}\cmidrule(lr){8-8} \cmidrule(lr){9-9} 
\multirow{4}{*}{{Our Method}} &Initial Search \tnote{*}
							& Ground Truth Vessel Maps  &0.25& {0.84}& {0.38} &0.26& {0.88}&0.40\\
						
	&\cellcolor{Gray} Refined Search\tnote{*} & \cellcolor{Gray}Ground Truth Vessel Maps  &\cellcolor{Gray}0.63& \cellcolor{Gray} \textbf{0.77}&\cellcolor{Gray} \textbf{0.69} &\cellcolor{Gray}\textbf{0.74}& \cellcolor{Gray} \textbf{0.82}&\cellcolor{Gray} \textbf{0.78}\\
&Initial Search	&  Fundus Images   &0.29&{0.85}& 0.43& 0.17& {0.87}&0.29 \\ 
&\cellcolor{Gray} Refined Search	  & \cellcolor{Gray}Fundus Images   &\cellcolor{Gray}0.65& \cellcolor{Gray}0.69& \cellcolor{Gray}0.67 &\cellcolor{Gray}0.52&\cellcolor{Gray}{0.67}&\cellcolor{Gray}0.59 \\ \midrule
&Pratt \textit{et al.} \cite{pratt2017automatic} \tnote{**} & Ground Truth Vessel Maps& 0.74&0.57& 0.64& 0.52&0.54& 0.53 \\
&\multirow{2}{*}{BICROS \cite{abbasi2016automatic}}&
 Fundus images $\&$Vessel Maps segmented with \cite{soares2006retinal} & \textbf{0.75}&0.61&{0.67}& 0.47&0.60&0.52\\
&&  Fundus images $\&$ Vessel Maps segmented with \cite{abbasi2015biologically} & \textbf{0.75}&0.61&0.66& {0.67}&0.57&{0.61}\\

\bottomrule
\end{tabular}
\begin{tablenotes}
\item[*] Junctions labeled by \cite{abbasi2016automatic} but not corresponding to any junctions in the ground truth vessel maps of the DRIVE were ignored during the performance evaluation. No refinement were performed for the junction labels in IOSTAR.
  \item[**] The result for the DRIVE belongs to the experiment where the network was trained on GRADE $1$ and tested on $G1A2$ \cite{abbasi2016automatic}.
 \end{tablenotes}
\end{threeparttable}
}
\end{table}

According to the table, recall rates at the initial search for both datasets are over $0.84$ regardless of image type (e.g. ground truth or fundus image), which indicates that junction probability maps have a high potential for the detection of almost $90\%$ of junctions in the ground truth data. However, precision at this stage is low because of the large number of false positives generated with the eigen-analysis. The refined search eliminates many of these false positives, with precision values increasing from $0.26$ up to $0.74$ (in the IOSTAR dataset) when ground truth images are input, and from $0.29$ up to $0.65$ (in the DRIVE dataset) when fundus images are input. Due to the elimination of some true positives at this stage, recall rates can drop for both image types; however, this is observed to be less recognizable when ground truth images are input. 

Considering input image types, the performance of the network is found to be better at ground truth images, particularly for IOSTAR with a $0.19$ increase on $F_{1}$ score for fundus images. However, we did not observe significant performance difference between the DRIVE and IOSTAR cases, which indicates the generalization ability of our method for unseen datasets even if they are captured by different imaging modalities.

Regarding the usage of a deep network, the most relevant study to ours is Pratt \textit{et al.}'s \cite{pratt2017automatic}, where a deep network was trained with vessel skeletons produced from ground truth vessel maps as a classifier to identify image patches carrying any types of junctions; then, local maxima in selected patches were labelled as junction locations. Although their tolerance distance to actual junctions, $10$ pixels, is two times larger than ours, our recall and $F_{1}$ score on the DRIVE and all the three metrics on IOSTAR are significantly larger than their findings. Moreover, the performance we obtained on fundus images for the same metrics, particularly for recall, was found to be still better than their performance on both datasets. 

BICROS \cite{abbasi2016automatic} combined two approaches: one relying on orientation scores obtained from fundus images and the other finding branches on skeletons generated from segmented vessel maps. Because of using different segmentation methods, their performance varies depending on suitability of segmentation methods to imaging modality \cite{abbasi2016automatic}. However, the approach we propose yielded better recall rates than theirs, regardless of imaging modalities and input image types.

\begin{figure}[!ht]
  \centering
  {\includegraphics[scale=0.7,keepaspectratio]{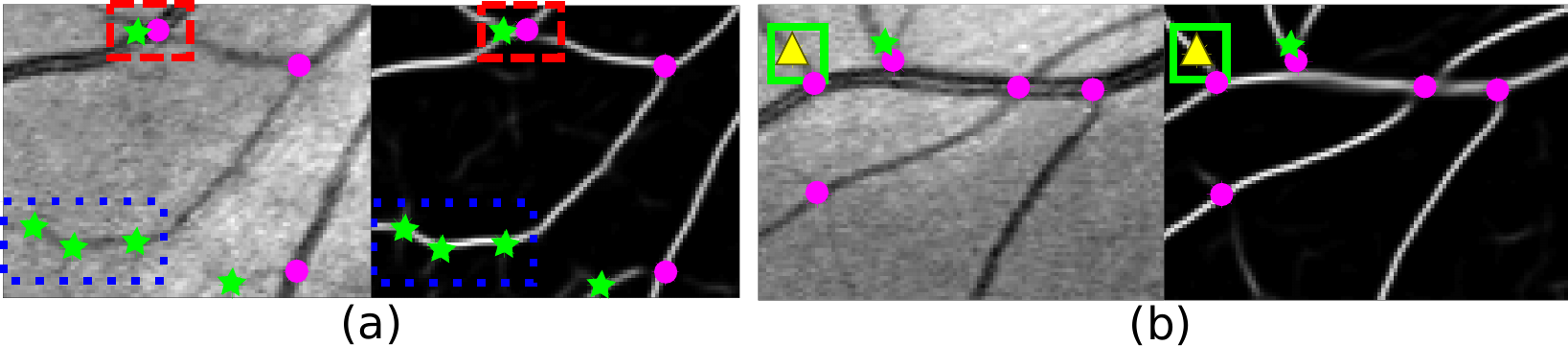}} %-eps-converted-to.pdf
  \vskip 0.2cm
\caption{Example results from our junction detection on a fundus image in the DRIVE, where each image pair contains a fundus image patch and its centerline probability map from left to right. Pink discs, green stars and yellow rectangles respectively show correctly detected, mistakenly detected and missing junctions. See the text for an explanation of the red dashed line, blue dotted line and green solid line rectangles. \label{fig:DRIVE19_Junctions}} 
\end{figure}

Precision rates reported by Pratt \textit{et al.} \cite{pratt2017automatic} and obtained with BICROS \cite{abbasi2016automatic} appear to be larger, particularly for the DRIVE, than ours. However, it should be noted that our method is not designed to differentiate crossings from bifurcations, and errors can occur if crossing vessels have large width \cite{calvo2011automatic}. This situation can be observed inside the red dashed line rectangle in Fig. \ref{fig:DRIVE19_Junctions}(a), where a crossing is represented with two joints. Also, some unlabeled junctions in ground truth data can reduce precision. False positives inside the blue dotted line rectangle in Fig. \ref{fig:DRIVE19_Junctions}(a) seem to be good candidates for junctions.  On the other hand, our method can fail to detect a few junctions if they are not represented with sufficient probabilities in centerline probability maps. The green solid line square in Fig. \ref{fig:DRIVE19_Junctions}(b) shows a junction missed by our method; but it is not easily seen by the naked eye.
\vspace{-0.3cm}
 
\section{Conclusion}
\vspace{-0.3cm}
Although deep networks have recently become the method of choice for medical image segmentation, their application is limited to those with large amounts of labeled data for the desired task. In order to deal with the scarcity of labeled data for junction detection in fundus images as a prior step to bifurcation and crossing classification, we propose a multi-task deep network. This network can produce probability maps of vessel interior, centerline, edges and junctions. Because of learning various descriptions of retinal vasculature, the network was found to confidently indicate junction locations. Potential junction locations suggested by the junction probability maps were reassessed by simply counting branch numbers on centerline probability maps. 

We evaluated the performance of our network on the DRIVE test set and IOSTAR after training the network with the DRIVE training set once. We found that the proposed approach outperformed previous approaches for the junction detection task. Moreover, the performance of the proposed approach appeared to be better than that of a similar method due to Pratt \textit{et al.} \cite{pratt2017automatic}, which used a popular deep network to identify junctions. Our findings suggest that the proposed method can be used for unseen retinal datasets, even if they have slightly different characteristics, without retraining the network.

\bibliography{main}

\end{document}